\documentclass[prb,twocolumn,showpacs]{revtex4}

\usepackage{bm}
\usepackage{amsmath}
\usepackage{amssymb}
\usepackage{graphicx}
\usepackage{amsfonts}
\usepackage{txfonts}
\usepackage{graphicx}

\begin{document}

\title{Spin states in InAs/AlSb/GaSb semiconductor quantum wells}
\author{Jun Li}
\affiliation{SKLSM, Institute of Semiconductors, Chinese Academy of Sciences, P.O. Box
912, Beijing 100083, China}
\author{Wen Yang}
\affiliation{SKLSM, Institute of Semiconductors, Chinese Academy of Sciences, P.O. Box
912, Beijing 100083, China}
\author{Kai Chang}
\email{kchang@red.semi.ac.cn}
\affiliation{SKLSM, Institute of Semiconductors, Chinese Academy of Sciences, P.O. Box
912, Beijing 100083, China}
\date{\today }

\begin{abstract}
We investigate theoretically the spin states in InAs/AlSb/GaSb broken-gap
quantum wells by solving the Kane model and the Poisson equation
self-consistently. The spin states in InAs/AlSb/GaSb quantum wells are quite
different from those obtained by the single-band Rashba model due to the
electron-hole hybridization. The Rashba spin-splitting of the lowest
conduction subband shows an oscillating behavior. The D'yakonov-Perel' spin
relaxation time shows several peaks with increasing the Fermi wavevector. By
inserting an AlSb barrier between the InAs and GaSb layers, the
hybridization can be greatly reduced. Consequently, the spin orientation,
the spin splitting, and the D'yakonov-Perel' spin relaxation time can be
tuned significantly by changing the thickness of the AlSb barrier.
\end{abstract}

\maketitle

\section{Introduction}

Heterostructures based on InAs, GaSb, and AlSb are one of the most promising
systems for fundamental physics research and design of novel devices due to
their advantages of high electron mobility, narrow band gap, strong
spin-obit coupling, and more importantly, the special broken-gap band lineup
at the InAs/GaSb interface\cite{Esaki,Sb_review}. In the past decades, the
properties of InAs/GaSb broken-gap supperlattices and quantum wells (QWs)
were investigated both theocratically and experimentally\cite%
{Band1,Band2,Band3,Band4,Band5,Band6,Band7,Band8,MiniGap1,MiniGap2,MiniGap3,MiniGap4}%
. These studies show that due to the overlap of InAs conduction band and
GaSb valence band, the energy dispersion may exhibits an anticrossing
behavior at a finite in-plane wave vector $\boldsymbol{k}_{\parallel }\neq 0$%
\cite{Band1,Band2,Band3,Band4,Band5,Band6,Band7,Band8}. The tunneling
between InAs and GaSb layers opens a mini hybridization gap, which was
observed experimentally\cite{MiniGap1,MiniGap2,MiniGap3,MiniGap4}. In
addition, the electrons in GaSb can move across InAs/GaSb interface into
InAs layer, forming a two-dimensional electron gas in InAs side and a
two-dimensional hole gas in GaSb side, which is promising for observing the
Bose-Einstein condensation of excitons\cite{BEC1,BEC2,MiniGap4}. In
practical applications, there have been many proposals for electronic and
optical devices utilizing the unique characteristics of InAs/AlSb/GaSb
system such as resonant tunneling structures\cite{RTD, RTD2}, infrared
detectors\cite{IFR}, and interband cascade laser diodes\cite{CLD}. Recently,
the spin-related properties of InAs/AlSb/GaSb system also attracted much
interest. For instance, there have been a number of spintronic device
proposals, including the Rashba spin filter\cite{Spinfilter}, the spin field
effect transistor\cite{Datta}, and the high-frequency optical modulator
utilizing the spin precession\cite{Optialmodulator}. Moreover, InAs/GaSb QW,
like HgTe/HgCdTe QW, is another possible candidate to demonstrate the
intrinsic spin Hall effect\cite{SpinHall} and quantum spin Hall phase\cite%
{QSpinHall} due to the inverted band structure.

In the previous works, the electron-hole hybridization in InAs/GaSb
broken-gap QWs have been well studied \cite%
{Band7,Band8,mixeffect1,mixeffect2}. It was shown that the \textquotedblleft
spin-up\textquotedblright~and the \textquotedblleft
spin-down\textquotedblright~states are affected differently by the
hybridization\cite{Band7,Band8}. This implies that the spin states near the
hybridization gap (where strong hybridization occurs) in InAs/GaSb QWs
should be quite different from those in conventional semiconductor QWs.
Since the Fermi level of the undoped InAs/GaSb QW lies inside the
hybridization gap\cite{EFproof}, the unusual spin states nearby would be
very important for many electronic properties.

In this paper, we investigate theoretically the spin orientation (i.e., the
expectation value of the Pauli operator $\boldsymbol{\sigma }\equiv 2%
\boldsymbol{S}$ for the electron spin $\boldsymbol{S}$ in an eigenstate),
the zero-field spin splitting, and the D'yakonov-Perel' (DP) spin relaxation%
\cite{SpinRelaxation} in InAs/AlSb/GaSb QWs based on the Kane model. The
charge transfer induced internal electric field is taken into account by
solving the Kane model and the Poisson equation self-consistently. For
InAs/AlSb/GaSb QWs, the spin orientation (or spin states) are significantly
modified by the hybridization between the conduction band and valence band,
which can be tuned by changing the thickness of AlSb barrier. The spin
orientation can be measured indirectly from the Faraday rotation, and indeed
offers us a physical picture and can lead to some novel effects such as the
persistent Spin Helix\cite{SpinHelix1}. Using the grating technique and
Faraday rotation, the persistent rotation of the spin orientation, i.e.,
spin helix of electron was observed in GaAs QW\cite{SpinHelix2}. The Rashba
spin-splitting (RSS) exhibits oscillating features as a function of in-plane
wave vector in InAs/AlSb/GaSb QW \cite{AnomalousSOI} near the hybridization
gap. The spin relaxation time, which are obtained from the perturbation
theory\cite{Averkiev}, shows that the unusual spin-splitting could lead to
several peaks with increasing the Fermi wave vector. These unusual features
all come from the strong electron-hole hybridization beyond the single band
model with linear Rashba spin-orbit interaction\cite{Rashba}. Interestingly,
the hybridization is very sensitive to the thickness of the AlSb barrier,
evidenced by a rapidly decreasing hybridization gap with increasing AlSb
barrier thickness. Consequently, all the spin-related properties, including
the spin orientation, the spin splitting, and the DP spin relaxation time,
can be tuned by varing the thickness of AlSb barrier. The property of
tunable spin states in InAs/AlSb/GaSb QWs might be useful in designing new
spintronic devices.

This paper is organized as follows. In Sec. \ref{sec:theory}, we describe
the theoretical method based on a self-consistent calculation combining the
Kane model and the Poisson equation. In Sec. \ref{sec:results}, we present
the numerical results for the band structure, the spin orientation, the
spin-splitting and DP spin relaxation time in InAs/AlSb/GaSb QWs. In Sec. IV
we give the conclusion.

\section{THEORY}

\label{sec:theory}

We consider an InAs/AlSb/GaSb broken-gap QW grown along the [001] direction
[see Fig. \ref{fig:fig1} (a)]. We choose the axis $x$, $y$, and $z$ to be
along [100], [010], and [001], respectively. Within the envelope function
approximation, the Kane model is a good starting point for systems with
strong electron-hole hybridization like InAs/AlSb/GaSb QWs. As InAs
conduction band overlaps with GaSb valence band, electrons could transfer
from GaSb layer into InAs layer. The charge redistribution induces an
internal electric field, which can be evaluated from the Poisson equation.
Generally, one needs to know the charge density distribution to solve the
Poisson equation, and the charge density distribution is in turn determined
by the electron wave function. Therefore a self-consistent procedure is
needed to take the charge-transfer effect into account in this system\cite%
{JPCSC}.

In Sec. \ref{sec:theoryA}, we discuss the Kane model and the self-consistent
calculation method. To consider the spin states in broken-gap QWs, we shall
give an explicit definition of \textquotedblleft spin-up\textquotedblright\
and \textquotedblleft spin-down\textquotedblright\ states by classifying the
eigenstates of the Kane model. This is discussed in Sec. \ref{sec:theoryB}.

\subsection{Hamiltonian and calculation method}

\label{sec:theoryA} Following the new envelope function theory\cite{Burt},
the Kane model which describes the bulk zinc-blende semiconductors can be
generalized to describe heterostructures by ordering the momentum operators
with respect to material parameters. By choosing the following set of basis
functions,

\begin{subequations}
\begin{align}
\phi_{1} & =\left\vert \frac{1}{2},\frac{1}{2}\right\rangle =\left\vert
S\uparrow\right\rangle , \\
\phi_{2} & =\left\vert \frac{1}{2},-\frac{1}{2}\right\rangle =\left\vert
S\downarrow\right\rangle , \\
\phi_{3} & =\left\vert \frac{3}{2},\frac{3}{2}\right\rangle =\frac{1}{\sqrt{2%
}}\left\vert (X+iY)\uparrow\right\rangle , \\
\phi_{4} & =\left\vert \frac{3}{2},\frac{1}{2}\right\rangle =\frac{i}{\sqrt{6%
}}\left\vert (X+iY)\downarrow-2Z\uparrow\right\rangle , \\
\phi_{5} & =\left\vert \frac{3}{2},-\frac{1}{2}\right\rangle =\frac{1}{\sqrt{%
6}}\left\vert (X-iY)\uparrow+2Z\downarrow\right\rangle , \\
\phi_{6} & =\left\vert \frac{3}{2},-\frac{3}{2}\right\rangle =\frac{i}{\sqrt{%
2}}\left\vert X-iY\downarrow\right\rangle , \\
\phi_{7} & =\left\vert \frac{1}{2},\frac{1}{2}\right\rangle =\frac{1}{\sqrt{3%
}}\left\vert (X+iY)\downarrow+Z\uparrow\right\rangle , \\
\phi_{8} & =\left\vert \frac{1}{2},-\frac{1}{2}\right\rangle =-\frac {i}{%
\sqrt{3}}\left\vert (X-iY)\uparrow-Z\downarrow\right\rangle ,
\label{eqn:basis}
\end{align}

the Kane Hamiltonian for zinc-blende crystals near the $\Gamma$ point is
\begin{widetext}
\begin{equation}
H_{k}=
\begin{bmatrix}
A & 0 & i\sqrt{3}V^{\dagger} & \sqrt{2}U & iV & 0 & iU & \sqrt{2}V\\
0 & A & 0 & -V^{\dagger} & i\sqrt{2}U & -\sqrt{3}V &
i\sqrt{2}V^{\dagger} &
-U\\
-i\sqrt{3}V & 0 & -(P+Q) & L & M & 0 & \frac{i}{\sqrt{2}}L & -i\sqrt{2}M\\
\sqrt{2}U & -V & L^{\dag} & -(P-Q) & 0 & M & i\sqrt{2}Q & i\sqrt{\frac{3}{2}%
}L\\
-iV^{\dag} & -i\sqrt{2}U & M^{\dag} & 0 & -(P-Q) & -L & -i\sqrt{\frac{3}{2}%
}L^{\dag} & i\sqrt{2}Q\\
0 & -\sqrt{3}V^{\dag} & 0 & M^{\dag} & -L^{\dag} & -(P+Q) &
-i\sqrt{2}M^{\dag}
& -\frac{i}{\sqrt{2}}L^{\dag}\\
-iU & -i\sqrt{2}V & -\frac{i}{\sqrt{2}}L^{\dag} & -i\sqrt{2}Q &
i\sqrt
{\frac{3}{2}}L & i\sqrt{2}M & -P-\Delta & 0\\
\sqrt{2}V^{\dag} & -U & i\sqrt{2}M^{\dag} &
-i\sqrt{\frac{3}{2}}L^{\dag} & -i\sqrt{2}Q & \frac{i}{\sqrt{2}}L & 0
& -P-\Delta
\end{bmatrix}
\label{eqn:Hamil}
\end{equation}
\end{widetext}
where
\end{subequations}
\begin{subequations}
\label{eq:Helements}
\begin{align}
A & =E_{v}+E_{g}+\boldsymbol{k}A_{c}\boldsymbol{k}, \\
P & =-E_{v}+\frac{\hbar^{2}}{2m_{0}}\boldsymbol{k}\gamma_{1}\boldsymbol{k},
\\
Q & =\frac{\hbar^{2}}{2m_{0}}(k_{x}\gamma_{2}k_{x}+k_{y}%
\gamma_{2}k_{y}-2k_{z}\gamma_{2}k_{z}), \\
L & =i\frac{\sqrt{3}\hbar^{2}}{m_{0}}\{k_{-}\gamma_{3}k_{z}\}, \\
M & =-\frac{\sqrt{3}\hbar^{2}}{2m_{0}}[k_{x}\gamma_{2}k_{x}-k_{y}\gamma
_{2}k_{y}-2i\{k_{x}\gamma_{3}k_{y}\}], \\
U & =\frac{1}{\sqrt{3}}P_{0}k_{z}, \\
V & =\frac{1}{\sqrt{6}}P_{0}k_{-}.\phantom{\gamma_{2}k_{x}+k_{y}\gamma_{2}
k_{y}-2k_{z}\gamma_{2}k_{z})}
\end{align}

In Eq. (\ref{eq:Helements}), $\boldsymbol{k}=(\boldsymbol{k}_{\parallel
},-i\partial/\partial_{z})$, $k_{\pm}=k_{x}\pm ik_{y}$, and $\{k_{\alpha
}\gamma k_{\beta}\}=\left( k_{\alpha}\gamma k_{\beta}+k_{\beta}\gamma
k_{\alpha}\right) /2$ $\left( \alpha,\beta=x,y,z\right) $. Here the in-plane
momentum as a constant of motion has been replaced by its eigenvalue $%
\boldsymbol{k}_{\parallel}$ and we have neglected the bulk inversion
asymmetry (Dresselhaus effect) since it is small compared to the structure
inversion asymmetry (Rashba effect). The strain effect is also neglected
because the lattice mismatch between InAs and GaSb is less than 1\%.

The band parameters are assumed to be a piecewise function along the growth
direction,
\end{subequations}
\begin{equation}
\gamma\left( z\right) =\sum_{i}\gamma^{i}\left[ \Theta\left( z-z_{i}\right)
-\Theta\left( z-z_{i+1}\right) \right] ,  \label{eq:gamma}
\end{equation}
where $\Theta\left( z\right) $ is the Heaviside step function, $z_{i}$ is
the $i$-th interface of this system, and $\gamma^{i}$ is the bulk band
parameters of the $i$-th layer. These band parameters, including $E_{g},\
E_{v},$ $\Delta,\ A_{c},\ \gamma_{1},\ \gamma_{2}$, and$\ \gamma _{3},$ can
be derived from the Luttinger parameters and the electron effective mass\cite%
{Parameters}. They are given in table \ref{tab:para}. The valence band
offset of two neighboring materials is $\Delta
E_{v}^{i}=E_{v}^{i}-E_{v}^{i+1}$.
\begin{table}[ptb]
\centering
\begin{ruledtabular}%
\caption{The Kane parameters used in our calculation. These
parameters are calculated from the Luttinger parameters obtained
from Ref.~\onlinecite{Parameters} (see the formulism below Table
\ref{tab:para}). The value of Kane energy
$E_{p}=2m_{0}P_{0}^{2}/\hbar^{2}$ is taken equal to 22.5 eV for each
layer material.}%
\begin{tabular}
[c]{llll} & InAs & GaSb & AlSb\\ \hline
$E_{g}$ (eV) & 0.417 & 0.812 & 2.386\\
$E_{v}$ (eV) & -0.417 & 0.143 & -0.237\\
$\Delta$ (eV) & 0.39 & 0.76 & 0.676\\
$A_{c}$ (eV$\cdot$nm$^{2}$) & -0.26 & 0.09 & -0.06\\
$\gamma_{1}$ & 2.01 & 4.16 & 2.04\\
$\gamma_{2}$ & -0.49 & 8.18 & -0.38\\
$\gamma_{3}$ & 0.21 & 1.38 & 0.40 \\
$\varepsilon $ & 14.55 & 15.69 & 14.4
\end{tabular}
\label{tab:para}%
\footnotetext[1] {The relation between Kane parameters and Luttinger
parameters are $\gamma_{1} = \gamma_{1}^{L}-\frac{E_{p}}{3E_{g}}$,
$\gamma_{2} = \gamma_{2}^{L}-\frac{E_{p}}{6E_{g}}$, $\gamma_{1} =
\gamma_{3}^{L}-\frac{E_{p}}{6E_{g}}$. }
\footnotetext[2] {$A_{c}$ is
correlated to electron effective mass $m^{*}$ by $A_{c}
=\frac{\hbar^{2}}{2m^{*}}-\frac{E_{p}}{6m_{0}}(\frac{2}{E_{g}}+\frac{1}%
{E_{g}+\Delta})$.}
\end{ruledtabular}
\end{table}

For thick enough InAs and GaSb layers, the lowest conduction subband in InAs
layer overlaps with the highest valence subband in GaSb layer. Electrons
could transfer from GaSb to InAs, inducing an internal electrostatic
potential $V_{in}\left( z\right) $. Therefore, the total Hamiltonian becomes
$H(\boldsymbol{k}_{\parallel })=H_{k}(\boldsymbol{k}_{\parallel
})-eV_{in}\left( z\right) $. The subband dispersions and the corresponding
eigenstates are obtained from the Schr\"{o}dinger equation
\begin{equation}
H(\boldsymbol{k}_{\parallel })\left\vert \Psi _{s}(\boldsymbol{k}_{\parallel
})\right\rangle =E_{s}(\boldsymbol{k}_{\parallel })\left\vert \Psi _{s}(%
\boldsymbol{k}_{\parallel })\right\rangle ,  \label{eq:sch}
\end{equation}%
where $s$ is the index of the subband and $\left\vert \Psi _{s}(\boldsymbol{k%
}_{\parallel })\right\rangle =\exp (i\boldsymbol{k}_{\parallel }\boldsymbol{%
\cdot \rho })[\varphi _{1}^{s}(z),\varphi _{2}^{s}(z),...,\varphi
_{8}^{s}(z)]^{T}$ is the envelope function. To solve the Schr\"{o}dinger
equation, we expand $\varphi _{n}^{s}$ by a series of plane waves,%
\begin{equation}
\varphi _{n}^{s}(z)=\frac{1}{\sqrt{L}}\sum_{m=-N}^{N}c_{nm}^{s}\exp
(ik_{m}z),  \label{eq:pwex}
\end{equation}%
where $k_{m}=2m\pi /L$ and $L$ is the total length of the structure (in this
work, $L=2L_{AlSb}^{side}+L_{InAs}+L_{AlSb}^{middle}+L_{GaSb}$, and $%
L_{AlSb}^{side}=2L_{InAs}$). By choosing a moderate $N$, one can also avoid
the spurious solutions that may occur in the Kane model.\cite%
{SpuriousSolution} In our calculation, $N\approx 25$ is good enough to get
convergent results.

The internal electrostatic potential $V_{in}\left( z\right) $ is determined
by the Poisson equation

\begin{equation}
\frac{d}{dz}\varepsilon (z)\frac{d}{dz}V_{in}(z)=-\left[ \rho _{e}(z)+\rho
_{h}(z)\right] ,  \label{eq:poisson}
\end{equation}%
where $\rho _{e}(z)$ and $\rho _{h}(z)$ are, respectively, the charge
density due to electrons and holes and $\varepsilon (z)$ is the static
dielectric constant. $\rho _{e}(z)$ and $\rho _{h}(z)$ can be derived from
the envelope functions:
\begin{equation}
\rho _{e}(z)=-\frac{e}{(2\pi )^{2}}\sum_{s}\int \sum_{n=1,2}\left\vert
\varphi _{n}^{s}(z)\right\vert ^{2}f_{F}(E_{s})d\boldsymbol{k_{\parallel }},
\label{eq:echarge}
\end{equation}%
\begin{equation}
\rho _{h}(z)=\frac{e}{(2\pi )^{2}}\sum_{s}\int \sum_{n=3,4,\cdots
,8}\left\vert \varphi _{n}^{s}(z)\right\vert ^{2}[1-f_{F}(E_{s}))]d%
\boldsymbol{k_{\parallel },}  \label{eq:hcharge}
\end{equation}%
where $f_{F}(E_{s})$ is the Fermi distribution function. The summations $%
\sum_{n=1,2}$ and $\sum_{n=3,4,\cdots ,8}$ runs over the electron components
$\phi _{1},\phi _{2}$ and hole components $\phi _{3},\cdots ,$ $\phi _{8}$
respectively. The summation $\Sigma_{s}$ includes all the subbands which
show anticrossing behavior. For simplicity, we take T = 0 K and the axial
approximation\cite{Sham, SLChuang} in the self-consistent procedure. Since
we want to give a clear picture about the effect of the hybridization on the
spin states, we only consider the anticrossing occurs when the lowest
conduction subband meets the highest valence subband, which limits to the
two cases $L_{InAs}\leq $ 14 nm at $L_{GaSb}=$ 10 nm and $L_{GaSb}\leq $ 14
nm at $L_{InAs}=$ 10 nm.

The Fermi level $E_{F}$ is determined by the charge neutrality condition.
\begin{equation}
\int_{0}^{L}\left[ \rho_{e}(z)+\rho_{h}(z)\right] dz=0.  \label{eq:neu}
\end{equation}
The Fermi level $E_{F}$ obtained by Eqs. (\ref{eq:echarge})-(\ref{eq:neu})
locates closely above the mini gap\cite{EFproof,EFdenote}. Combining Eqs. (%
\ref{eq:sch})-(\ref{eq:neu}), we can do a self-consistent iteration that
eventually yields the internal electrostatic potential $V_{in}\left(
z\right) $. Once we have $V_{in}\left( z\right) $, the subband dispersions
and electronic states can be obtained by solving Eq. (\ref{eq:sch}).

\subsection{The classification of the spin states}

\label{sec:theoryB}

In the single-band Rashba model for the electron (i.e., $\phi_{1}$ and $%
\phi_{2}$ components only), there is a well-defined spin quantization axis $%
\hat{\boldsymbol{e}}_{\Sigma}=\hat{\boldsymbol{e}}_{\boldsymbol{k}%
_{\parallel}}\times\hat{\boldsymbol{e}}_{z}$ perpendicular to both the wave
vector direction $\hat{\boldsymbol{e}}_{\boldsymbol{k}_{\parallel}}$ and the
QW growth direction $\hat{\boldsymbol{e}}_{z}$. The spin orientation of any
eigenstate always has a magnitude of unity and is either parallel (called
\textquotedblleft spin-up\textquotedblright\ eigenstates) or anti-parallel
(called \textquotedblleft spin-down\textquotedblright\ eigenstates) to $\hat{%
\boldsymbol{e}}_{\Sigma}$\cite{ElectronSpin}. For the $J=3/2$ hole system
(i.e., components $\phi_{3},\cdots,\phi_{6}$) or a hybridized electron-hole
system\cite{HoleSpin}, however, such a quantization axis does not exist. As
we shall show in Sec. \ref{sec:results}, for a given conduction subband, the
spin orientation may change its magnitude and direction (up to 180 degrees)
with increasing $\boldsymbol{k}_{\parallel}$. The absence of a well-defined
spin quantization axis make it impossible to classify the \textquotedblleft
spin-up\textquotedblright~and \textquotedblleft
spin-down\textquotedblright~states by their spin orientations relative to
this quantization axis.

In the following, we divide the Kane Hamiltonian $H_{k}(\boldsymbol{k}%
_{\parallel})$ in Eq. (\ref{eqn:Hamil}) (note that bulk inversion asymmetry
has been neglected) as the sum of a dominanting part $H_{ax}(\boldsymbol{k}%
_{\parallel})$ with axial symmetry and a small cubic part $H_{cub}(%
\boldsymbol{k}_{\parallel})$ with cubic symmetry\cite{SOIRWinkler}
\begin{equation}
H_{k}(\boldsymbol{k}_{\parallel})=H_{ax}(\boldsymbol{k}_{\parallel})+H_{cub}(%
\boldsymbol{k}_{\parallel}).  \label{eq:Hdecompose}
\end{equation}
The axial part $H_{ax}(\boldsymbol{k}_{\parallel})$ obtained from Eq. (\ref%
{eqn:Hamil}) by replacing $\gamma_{2}$ and $\gamma_{3}$ with $\overline{%
\gamma}$ in the term $M$ and the cubic part $H_{cub}(\boldsymbol{k}%
_{\parallel})\equiv H_{k}(\boldsymbol{k}_{\parallel})-H_{ax}(\boldsymbol{k}%
_{\parallel})$ is the difference between $H_{k}(\boldsymbol{k}_{\parallel})$
and $H_{ax}(\boldsymbol{k}_{\parallel})$.

The axial part $H_{ax}(\boldsymbol{k}_{\parallel})$ can be transformed into
a block-diagonal form by choosing a new basis set. Similar transformation
has been reported in dealing with the four-band\cite{Sham} and six-band
models\cite{SLChuang}, but the cubic part was neglected in these works. In
this paper the transformation is extended to the eight-band model and the
cubic part is also included. Let $\boldsymbol{k}_{\parallel
}=k_{\parallel}(\cos\varphi,\sin\varphi)$ and the new basis set is $\varphi$%
-dependent,
\begin{align}
\left\vert S(-)\right\rangle & =\frac{1}{\sqrt{2}}(i\phi_{1}-e^{i\varphi
}\phi_{2}), & \left\vert S(+)\right\rangle & =\frac{1}{\sqrt{2}}(i\phi
_{1}+e^{i\varphi}\phi_{2}),  \notag  \label{eqn:newbasis} \\
\left\vert HH(-)\right\rangle & =\frac{e^{-i\varphi}}{\sqrt{2}}(\phi
_{3}+e^{3i\varphi}\phi_{6}), & \left\vert HH(+)\right\rangle & =\frac{%
e^{-i\varphi}}{\sqrt{2}}(\phi_{3}-e^{3i\varphi}\phi_{6}),  \notag \\
\left\vert LH(-)\right\rangle & =\frac{1}{\sqrt{2}}(\phi_{4}+e^{i\varphi
}\phi_{5}), & \left\vert LH(+)\right\rangle & =\frac{1}{\sqrt{2}}(\phi
_{4}-e^{i\varphi}\phi_{5}),  \notag \\
\left\vert SO(-)\right\rangle & =\frac{i}{\sqrt{2}}(\phi_{7}+e^{i\varphi
}\phi_{8}), & \left\vert SO(+)\right\rangle & =\frac{i}{\sqrt{2}}(\phi
_{7}-e^{i\varphi}\phi_{8}),
\end{align}
Under this basis, $H_{ax}(\boldsymbol{k}_{\parallel})\rightarrow \mathcal{H}%
_{ax}(k_{\parallel})$ is block-diagonalized and $H_{cub}(\boldsymbol{k}%
_{\parallel})\rightarrow\mathcal{H}_{cub}(\boldsymbol{k}_{\parallel})$ has a
simple dependence on the azimuth $\varphi$,

\begin{align}
\mathcal{H}_{ax}(k_{\parallel}) & =%
\begin{pmatrix}
\mathcal{H}_{-}(k_{\parallel}) & 0 \\
0 & \mathcal{H}_{+}(k_{\parallel})%
\end{pmatrix}
, \\
\mathcal{H}_{cub}(\boldsymbol{k}_{\parallel}) & =%
\begin{pmatrix}
\mathcal{A}(k_{\parallel})\cos4\varphi & -i\mathcal{A}(k_{\parallel})\sin4%
\varphi \\
i\mathcal{A}(k_{\parallel})\sin4\varphi & -\mathcal{A}(k_{\parallel})\cos4%
\varphi%
\end{pmatrix}
,  \label{eqn:HaxHcubB}
\end{align}
where

\begin{equation}
\mathcal{H}_{\pm}(k_{\parallel})=%
\begin{bmatrix}
A & \sqrt{3}V^{\prime} & -\sqrt{2}U^{\prime}\mp V^{\prime} & U^{\prime}\mp%
\sqrt{2}V^{\prime} \\
\sqrt{3}V^{\prime} & -P-Q & L^{\prime}\mp M^{\prime} & -\frac{%
L^{\prime}\pm2M^{\prime}}{\sqrt{2}} \\
\sqrt{2}U^{\prime}\mp V^{\prime} & -L^{\prime}\mp M^{\prime} & -P+Q & \frac{%
-2Q\pm\sqrt{3}L^{\prime}}{\sqrt{2}} \\
-U^{\prime}\mp\sqrt{2}V^{\prime} & \frac{L^{\prime}\mp2M^{\prime}}{\sqrt{2}}
& \frac{-2Q\mp\sqrt{3}L^{\prime}}{\sqrt{2}} & -P-\Delta%
\end{bmatrix}
,  \label{eqn:Hpm}
\end{equation}

\begin{equation}
\mathcal{A}(k_{\parallel}) = \frac{\sqrt{3}\hbar^{2}k_{\parallel}^{2}}{2m_{0}%
}\Delta\overline{\gamma}
\begin{bmatrix}
0 & 0 & 0 & 0 \\
0 & 0 & 1 & \sqrt{2} \\
0 & 1 & 0 & 0 \\
0 & \sqrt{2} & 0 & 0%
\end{bmatrix}
,  \label{eqn:AMatrix}
\end{equation}
and

\begin{subequations}
\begin{align}  \label{eq:LMUV2}
L^{\prime} & =i\frac{\sqrt{3}\hbar^{2}}{2m_{0}}k_{\parallel
}(\gamma_{3}k_{z}+k_{z}\gamma_{3}), \\
M^{\prime} & =-\frac{\sqrt{3}\hbar^{2}}{2m_{0}}\overline{\gamma}%
k_{\parallel}^{2}, \\
U^{\prime} & =\frac{i}{\sqrt{3}}P_{0}k_{z}, \\
V^{\prime} & =\frac{1}{\sqrt{6}}P_{0}k_{\parallel}, \\
\overline{\gamma} & =\frac{1}{2}(\gamma_{2}+\gamma_{3}), \\
\Delta\overline{\gamma} & =\frac{1}{2}(\gamma_{2}-\gamma_{3}).
\end{align}

Using the new basis set, we can give an explicit classification of all the
eigenstates into \textquotedblleft spin-up\textquotedblright\ and
\textquotedblleft spin-down\textquotedblright\ states, similar to the
single-band model.

In the absence of the cubic term $\mathcal{H}_{cub}(\boldsymbol{k}%
_{\parallel })$ (so-called axial approximation), the eigenstates $%
|\Psi_{s,\pm}^{(ax)}(k_{\parallel})\rangle$ and eigen-energies $%
E_{s,\pm}^{(ax)}(k_{\parallel})$ (for the $s$-th subband) of the total
Hamiltonian $H(\boldsymbol{k}_{\parallel})=\mathcal{H}_{ax}(k_{%
\parallel})-eV_{in}\left( z\right) $ are determined by
\end{subequations}
\begin{equation}
\left[ \mathcal{H_{\pm}}(k_{\parallel})-eV_{in}(z)\right] \left\vert
\Psi_{s,\pm}^{(ax)}(k_{\parallel})\right\rangle
=E_{s,\pm}^{(ax)}(k_{\parallel})\left\vert
\Psi_{s,\pm}^{(ax)}(k_{\parallel})\right\rangle ,
\end{equation}
with the Rashba spin splitting in the $s$-th subband
\begin{equation}
\Delta
E_{s}^{(ax)}(k_{\parallel})=E_{s,+}^{(ax)}(k_{\parallel})-E_{s,-}^{(ax)}(k_{%
\parallel}).
\end{equation}
Obviously, the eigenstates are automatically classified into two classes:
the \textquotedblleft spin-down\textquotedblright~states $%
|\Psi_{s,-}^{(ax)}(k_{\parallel})\rangle$ consisting of the components $%
\left\vert S(-)\right\rangle $, $\left\vert HH(-)\right\rangle $, $%
\left\vert LH(-)\right\rangle $, and $\left\vert SO(-)\right\rangle $ and
the \textquotedblleft spin-up\textquotedblright~states $%
|\Psi_{s,+}^{(ax)}(k_{\parallel})\rangle$ consisting of the components $%
\left\vert S(+)\right\rangle $, $\left\vert HH(+)\right\rangle $, $%
\left\vert LH(+)\right\rangle $, and $\left\vert SO(+)\right\rangle $. Then
one can define the spin orientation of an arbitrary eigenstate $%
|\Psi_{s,\pm}^{(ax)}(k_{\parallel})\rangle$ as\cite{ElectronSpin,HoleSpin}
\begin{equation}
\langle\boldsymbol{\Sigma}^{(ax)}(\boldsymbol{k}_{\parallel})\rangle_{s,\pm
}=\left\langle \Psi_{s,\pm}^{(ax)}(k_{\parallel})\right\vert \boldsymbol{%
\Sigma}(\varphi)\left\vert \Psi_{s,\pm}^{(ax)}(k_{\parallel })\right\rangle ,
\end{equation}
where $\boldsymbol{\Sigma}(\varphi)\equiv\lbrack\Sigma_{x}(\varphi),\Sigma
_{y}(\varphi),\Sigma_{z}(\varphi)]$ are $8\times8$ matrices (see Appendix %
\ref{appen:SM}) for the Pauli operators $\boldsymbol{\sigma}\equiv 2%
\boldsymbol{S}$ in the new basis (\ref{eqn:newbasis}).

In the presence of the cubic part $\mathcal{H}_{cub}(\boldsymbol{k}%
_{\parallel})$, the off-block-diagonal terms $\pm i\mathcal{A}\sin4\varphi$
will in general couple the \textquotedblleft spin-up\textquotedblright\
states with \textquotedblleft spin-down\textquotedblright\ states, unless $%
\boldsymbol{k_{\parallel}}$ points along a high symmetry axis satisfying $%
\sin4\varphi=0$. As a result, the exact eigenstate $\left\vert \Psi _{s}(%
\boldsymbol{k_{\parallel}})\right\rangle $ of the Kane Hamiltonian is in
general a mixture of \textquotedblleft spin-up\textquotedblright\ and
\textquotedblleft spin-down\textquotedblright\ states. For convenience,
however, we still classify the exact eigenstates as \textquotedblleft
spin-up\textquotedblright\ or \textquotedblleft spin-down\textquotedblright\
states according to the dominant component.

Now we discuss the electronic structure of InAs/GaSb and InAs/AlSb/GaSb QWs
in the axial approximation [$\mathcal{H}_{cub}(\boldsymbol{k}_{\parallel})=0$%
] and the modification due to the cubic correction $\mathcal{H}_{cub}(%
\boldsymbol{k}_{\parallel})$. In the axial approximation, the electronic
structure shows three distinct features:

\begin{enumerate}
\item Electron-hole anticrossing. If the InAs and GaSb layers are thick
enough, the lowest conduction subband ($CB1$) anticrosses with the highest
valence subband ($VB1$) at a critical in-plane wave vector $\boldsymbol{k}%
_{a}$, evidenced by the switch of the dominant component (a) from $%
\left\vert HH(\pm)\right\rangle $ to $\left\vert S(\pm)\right\rangle $ for
the $CB1$ subband state $|\Psi_{CB1,\pm}^{(ax)}(k_{\parallel})\rangle$ and
(b) from $\left\vert S(\pm)\right\rangle $ to $\left\vert
HH(\pm)\right\rangle $ for the $VB1$ subband state $|\Psi_{VB1,%
\pm}^{(ax)}(k_{\parallel})\rangle$ when $\boldsymbol{k}_{\parallel}$ is
increased across $\boldsymbol{k}_{a}$.

\item The spin orientation of $|\Psi_{s,\pm}^{(ax)}(k_{\parallel})\rangle$
is always along $\hat{\boldsymbol{e}}_{\Sigma}=\hat{\boldsymbol{e}}_{%
\boldsymbol{k}_{\parallel}}\times\hat{\boldsymbol{e}}_{z}$, i.e., in the QW
plane and perpendicular to the in-plane wave vector $\boldsymbol{k}%
_{\parallel}$, in agreement with the prescription of the single-band Rashba
model\cite{Rashba, ElectronSpin} (this can be verified using Eqs. (\ref%
{eq:SOdn})-(\ref{eq:SOup}) in Appendix \ref{appen:SM}). But the projection
of the spin orientation varies with $\boldsymbol{k}_{\parallel}$ and may
even change sign, which is different from the single-band Rashba model.

\item The Rashba spin-splitting of the $CB1$ subband exhibits an oscillating
behavior [containing zero spin-splitting (spin degeneracy) points] due to
the electron-hole anticrossing.
\end{enumerate}

In the presence of the cubic correction $\mathcal{H}_{cub}(\boldsymbol{k}%
_{\parallel}),$ the following additional features are introduced:

\begin{enumerate}
\item The band structure and spin-splitting as a function of $\boldsymbol{k}%
_{\parallel}$ display a $C_{4v}$ symmetry.

\item The spin orientation of the exact eigenstate $\left\vert \Psi _{s}(%
\boldsymbol{k_{\parallel}})\right\rangle $ deviates from $\hat {\boldsymbol{e%
}}_{\Sigma}=\hat{\boldsymbol{e}}_{\boldsymbol{k}_{\parallel}}\times\hat{%
\boldsymbol{e}}_{z}$ (although it still lies in the QW plane) due to
hybridization of spin-up and spin-down states. The maximum deviation arises
when the azimuth angle $\varphi$ of $\boldsymbol{k}_{\parallel}$ satisfies $%
\sin(4\varphi)=\pm1$ [such that the spin mixing matrix element in Eq. (\ref%
{eqn:HaxHcubB}) reaches the maximum] and at one of the spin degeneracy
points.
\end{enumerate}

In Sec. \ref{sec:results}, all these special properties in InAs/GaSb and
InAs/AlSb/GaSb broken-gap QWs are demonstrated by our numerical results.

\section{Numerical Results and Discussion}

\label{sec:results}

\begin{figure}[h]
\includegraphics[width=1\columnwidth]{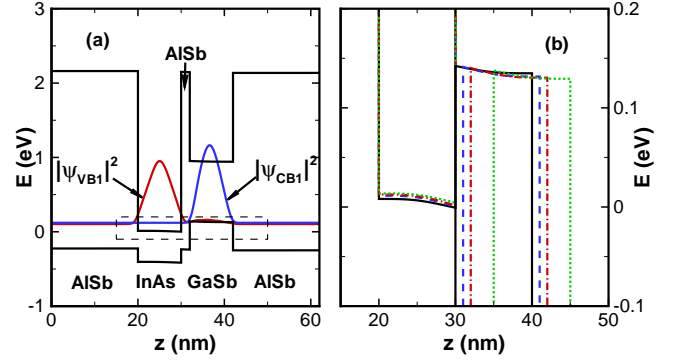}
\caption{(Colour online) (a) Band profile and the probability density
distribution of CB1($\pm $) (blue line), VB1($\pm $) (red line) states in a
10-2-10 nm InAs/AlSb/GaSb quantum well at $k_{\parallel}=0$; (b) the
amplification of the dashed square area in panel (a). The black solid, blue
dashed, red dashdot and green dotted line in panel (b) denotes the results
of $L_{AlSb}=$ 0 nm, 1 nm, 2 nm and 5 nm respectively. }
\label{fig:fig1}
\end{figure}

\begin{figure}[hh]
\includegraphics[width=1\columnwidth]{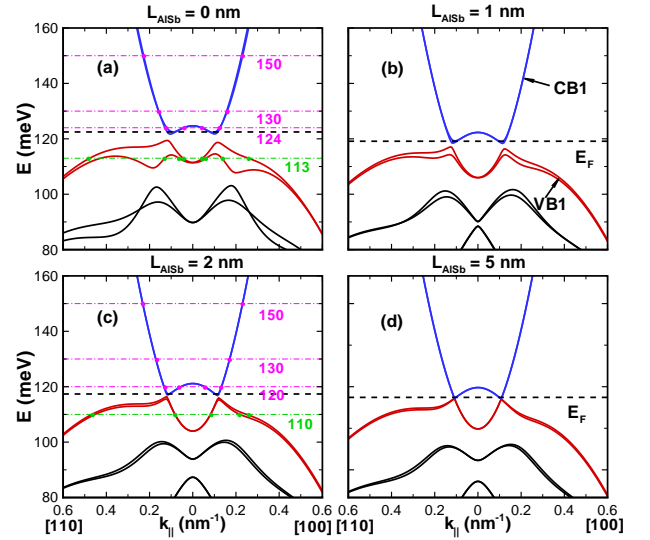}
\caption{(Colour online) The subband dispersions of InAs/AlSb/GaSb quantum
well structures with different thicknesses of AlSb layers: (a) $L_{AlSb}=0$
(b) $L_{AlSb}=1$ nm (c) $L_{AlSb}=2$ nm (d) $L_{AlSb}=5$ nm. The thicknesses
of InAs and GaSb layers are fixed at 10 nm. The Fermi energy denoted by the
dashed lines lies at the bottom of the lowest conduction band, in agreement
with previous works\protect\cite{Band1,EFproof}. In panel (a) and (c), the
cross points between CB1 (VB1) and the purple (green) dashdotted lines mark
the constant energy contours drawn in Fig. \protect\ref{fig:fig3} (Fig.
\protect\ref{fig:fig4}).}
\label{fig:fig2}
\end{figure}

In Fig. \ref{fig:fig1} we show the self-consistent band profile of an
InAs/AlSb/GaSb broken-gap QWs. In this QW, the concentration of electrons
transferred from GaSb layer to InAs layer is found to be of the order 1.61 $%
\times$ 10$^{11}$ cm$^{-2}$. The resulting electronstatic potential $%
V_{in}(z)$ induces a 8$\sim$14 meV downward (upward) bending of the InAs
conduction band (GaSb valence band) near the interface and causes a slight
shift of the subbands (see Fig. \ref{fig:fig1} (b) ). Note that we denote
the upper two branches of anticrossed subbands as $CB1(\pm)$ and the lower
two as $VB1(\pm)$ (see Fig. \ref{fig:fig2}), which are different from the
previous works\cite{Band7,Band8}.

In Fig. \ref{fig:fig2} we plot the band structure of InAs/AlSb/GaSb QWs with
different thicknesses of middle AlSb barrier (The thicknesses of InAs and
GaSb layers are fixed at 10 nm). Due to electron-hole hybridization, the $%
CB1 $ and $VB1$ subbands exhibit a strong anticrossing behavior and open a
mini hybridization gap at a finite $\boldsymbol{k}_{\parallel}$, consistent
with the previous works \cite{Band3,Band4,Band5,Band6,Band7,Band8}. The
resulting strongly-hybridized states near the gap may have significant
contributions to the spin-related properties of the broken-gap QWs, since
the Fermi energy (the dashed lines in Fig. \ref{fig:fig2}) locates nearby
\cite{EFproof,EFdenote}. From Fig. \ref{fig:fig2} (b)-(d), we can see that
by increasing the thickness of the AlSb barrier layer, the anticrossing
between $CB1$ and $VB1$ is gradually weakened and the hybridized gap is
noticeably narrowed due to the tunneling between $CB1$ and $VB1$ is
suppressed significantly. Meanwhile, the spin splitting of each subband is
greatly reduced due to the decreasing structural inversion asymmetry in the
InAs/AlSb/GaSb QW compared with the InAs/GaSb QW.

\begin{figure}[t]
\includegraphics[width=1\columnwidth]{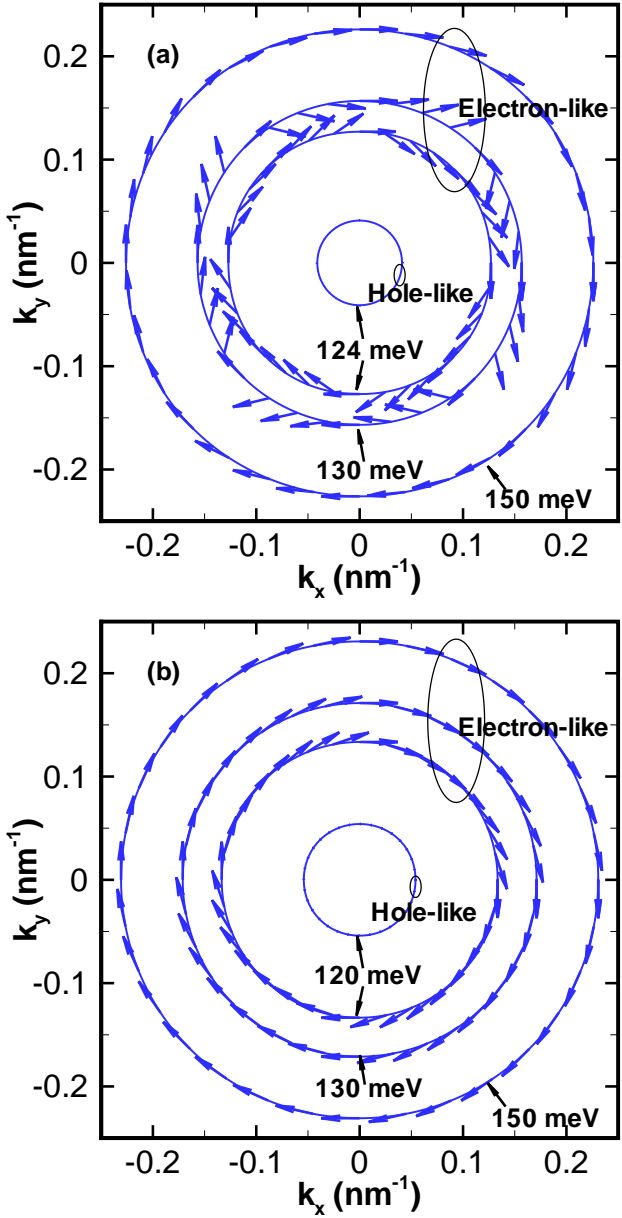} 
\caption{(Colour online) The spin orientations on the constant energy
contours of $CB1(+)$ subband of (a) a 10 nm InAs/10 nm GaSb QW; (b) a 10 nm
InAs/2 nm AlSb/10 nm GaSb QW. Note the dominant components of the states of
the insidest contours are $\left\vert HH(+)\right\rangle $, while that of
the other contours are $\left\vert S(+)\right\rangle $.}
\label{fig:fig3}
\end{figure}

\begin{figure}[t]
\includegraphics[width=1\columnwidth]{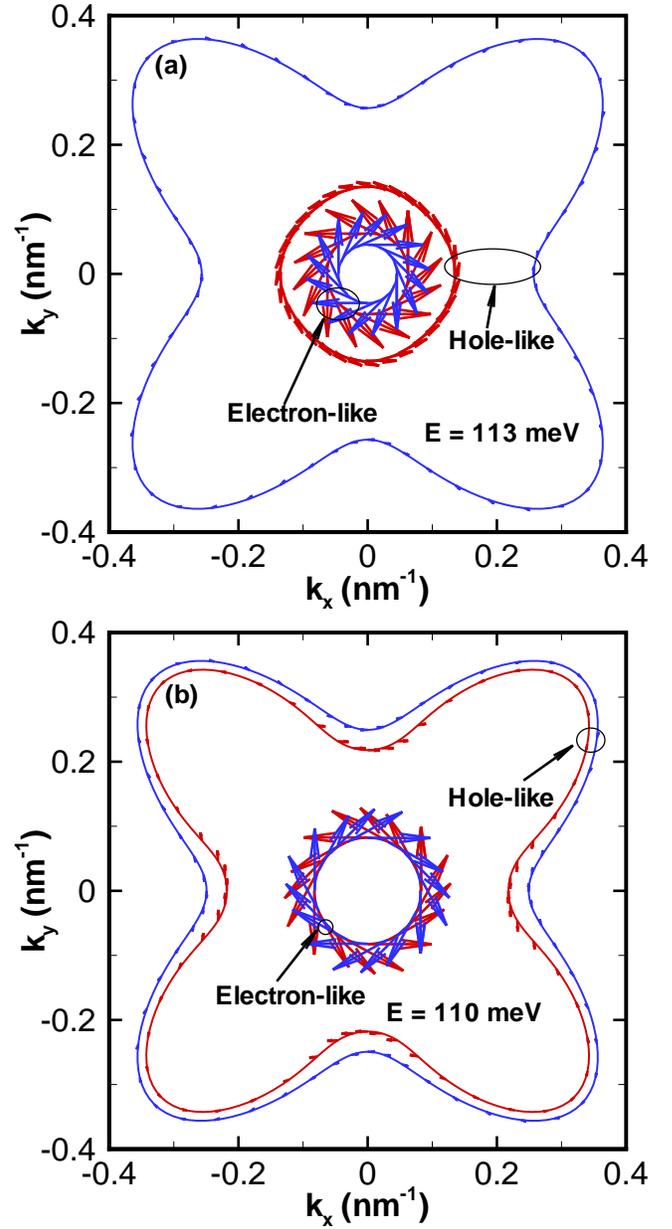} 
\caption{(Colour Online) The spin orientations on the constant energy
contours of $VB1$ subband of (a) a 10 nm InAs/10 nm GaSb QW; (b) a 10 nm
InAs/2 nm AlSb/10 nm GaSb QW. The blue and red lines represent $VB1(+)$ and $%
VB1(-)$ branches respectively.}
\label{fig:fig4}
\end{figure}

In Fig. \ref{fig:fig3} (a) we show the spin orientations of the eigenstates
on different constant energy contours of $CB1(+)$ subband in InAs/GaSb QW.
The spin orientations on the contour of $CB1(-)$ subband are antiparallel to
those of $CB1(+)$ subband and are omitted in the figure for brevity. In the
\textit{k}-linear Rashba model, the spin orientations of an eigenstate is
along $\hat{\boldsymbol{e}}_{\Sigma }=\hat{\boldsymbol{e}}_{\boldsymbol{k}%
_{\parallel }}\times \hat{\boldsymbol{e}}_{z}$, i.e., along the tangent
direction of the circular energy contour in the QW plane \cite{ElectronSpin}%
. However, this property no longer holds for InAs/AlSb/GaSb broken-gap QW.
The spin orientations deviate strongly from the tangent direction $\hat{%
\boldsymbol{e}}_{\boldsymbol{k}_{\parallel }}\times \hat{\boldsymbol{e}}_{z}$%
, unless $\boldsymbol{k_{\parallel }}$ points along high symmetry directions
(such as $\left\langle 100\right\rangle $ and $\left\langle 110\right\rangle
$) satisfying $\sin 4\varphi =0$. This comes from the hybridization between
the \textquotedblleft spin-up\textquotedblright ~and \textquotedblleft
spin-down\textquotedblright ~states, as discussed in the previous section.
Therefore we can see the spin orientations on the contour $E=130$ meV
deviate the most heavily because this contour is nearest to the maximum
hybridization point, and when $\boldsymbol{k_{\parallel }}$ lies in the
directions $\varphi =\pi /8,~3\pi /8$ ($\sin 4\varphi =\pm 1$), the maximum
hybridization occurs. When we insert an AlSb barrier between InAs and GaSb
layers, the hybridization is strongly reduced [Fig. \ref{fig:fig3}(b)]. Thus
the deviation of the spin orientation from the tangent direction is very
small. The results of Fig. \ref{fig:fig3} imply that one can tune the spin
orientations near the Fermi level by changing the thickness of AlSb barrier
in the middle of InAs and GaSb layers.

Fig. \ref{fig:fig4} (a) exhibits the spin orientations for the states on
different constant energy contours of $VB1$ subband in a InAs/GaSb QW. The
energy contours of $VB1$ subband show a very complicated behavior and a
strong anisotropy in $[100]$ and $[110]$ directions due to the complicated
band structures (see Fig. \ref{fig:fig2}). From the figure, one can easily
find a $C_{4v}$ group symmetry, which comes from cubic symmetry of the
crystal. For $E=113$ meV, we can find two pairs of contours. The states are
electron-like for the inner but hole-like for the outer pair. Due to the
large spin splitting, the shape of energy contour for \textquotedblleft
spin-up\textquotedblright ~and \textquotedblleft spin-down\textquotedblright
~state looks very different. If we insert an AlSb barrier between InAs and
GaSb layers, the spin-splitting between $VB1(+)$ and $VB1(-)$ is greatly
reduced (see Fig. (\ref{fig:fig4}) (b)).

\begin{figure}[t]
\includegraphics[width=1\columnwidth]{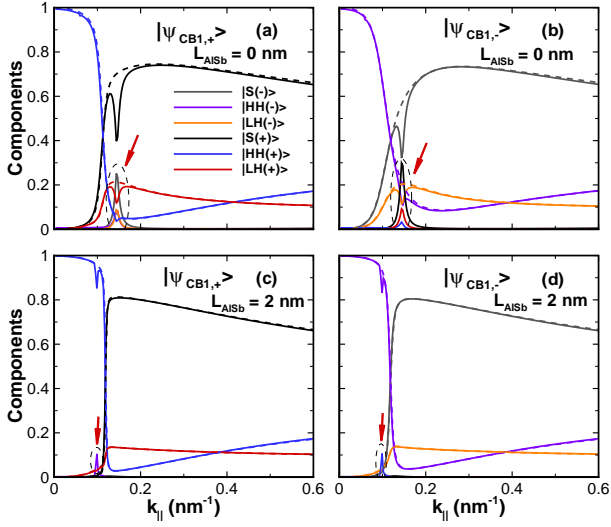} 
\caption{(Colour online) The components of states in $CB1(\pm)$ subband as a
function of $\boldsymbol{k_{\parallel}}$ along $\protect\varphi=\protect\pi%
/8 $ (solid line) and $\protect\varphi=\protect\pi/4$ ([110]) direction
(dashed line) in a 10 nm InAs/10 nm GaSb QW (a), (b) and a 10 nm InAs/2 nm
AlSb/10 nm GaSb QW (c), (d). The red arrows indicate the maximum
hybridization point. }
\label{fig:fig5}
\end{figure}

In order to demonstrate the hybridization of the \textquotedblleft
spin-up\textquotedblright ~and \textquotedblleft spin-down\textquotedblright
~state, we plot the components $\left\vert S(\pm )\right\rangle $, $%
\left\vert HH(\pm )\right\rangle $, $\left\vert LH(\pm )\right\rangle $ of
states in $CB1(\pm )$ subbands in Fig. \ref{fig:fig5}. From this figure, one
can see that the components $\left\vert HH(\pm )\right\rangle $ and $%
\left\vert S(\pm )\right\rangle $ varies significantly when increase $%
\boldsymbol{k}_{\parallel }$ over the anticrossing point $\boldsymbol{k}_{a}$%
\cite{Band7,Band8,AnomalousSOI}. Interestingly, at $\varphi =\pi /8$
direction, a strong hybridization between the \textquotedblleft
spin-up\textquotedblright ~state and \textquotedblleft
spin-down\textquotedblright ~state in the $CB1$ subbands occurs. This
feature can be proven by the peak of \textquotedblleft
spin-down\textquotedblright ~component and the dip of \textquotedblleft
spin-up\textquotedblright ~component at $\boldsymbol{k}_{\parallel
}=0.145~nm^{-1}$ in the $CB1(+)$ subband. A similar behavior appears in the $%
CB1(-)$) subband. For $\boldsymbol{k}_{\parallel }$ along [110] ($\sin
(4\varphi )=0$), there is no hybridization in the $CB1$ subbands and leading
to pure \textquotedblleft spin-up\textquotedblright ~and \textquotedblleft
spin-down\textquotedblright ~states . By inserting an AlSb barrier between
InAs and GaSb layers, the hole-like to electron-like transition of $%
\left\vert \Psi _{CB1,\pm }\right\rangle $ states become more emergent, and
the hybridization between \textquotedblleft spin-up\textquotedblright ~or
\textquotedblleft spin-down\textquotedblright ~state is reduced.

\begin{figure}[t]
\includegraphics[width=1\columnwidth]{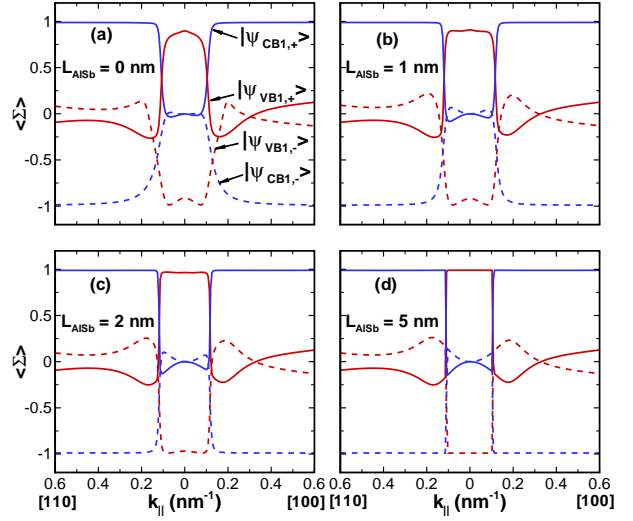}
\caption{(Colour online) The projection of spin expectation values of $%
\left| \Psi_{CB1,\pm}\right> $ and $\left| \Psi_{VB1,\pm}\right> $ states
along the orthogonal direction of $\boldsymbol{k_{\parallel}}$ in
InAs/AlSb/GaSb quantum well structures with different thicknesses of AlSb
layers: (a) $L_{AlSb} = 0$ (b) $L_{AlSb} = 1$ nm (c) $L_{AlSb} = 2$ nm (d) $%
L_{AlSb} = 5$ nm. The thickness of InAs and GaSb layers are fixed at 10 nm.}
\label{fig:fig6}
\end{figure}

As a result of the dominant component transition of the $\left\vert \Psi
_{CB1,\pm }\right\rangle $ and $\left\vert \Psi _{VB1,\pm }\right\rangle $
states when $\boldsymbol{k}_{\parallel }$ sweeps across the anticrossing
point $\boldsymbol{k}_{a}$, the spin expectation value magnitude $\langle
\Sigma \rangle $ change correspondingly. In Fig. (\ref{fig:fig6}) we display
the change of $\langle \Sigma \rangle $ for $\left\vert \Psi _{CB1,\pm
}\right\rangle $ and $\left\vert \Psi _{VB1,\pm }\right\rangle $ states as a
function of $\boldsymbol{k}_{\parallel }$. $\langle \Sigma \rangle $ can be
defined by projecting the vector $\langle \boldsymbol{\Sigma }\rangle $ onto
$\hat{\boldsymbol{e}}_{\Sigma }$, with $\hat{\boldsymbol{e}}_{\Sigma }=\hat{%
\boldsymbol{e}}_{\boldsymbol{k}_{\parallel }}\times \hat{\boldsymbol{e}}_{z}$
is the unit vector of the in-plane direction perpendicular to $\boldsymbol{%
k_{\parallel }}$. A sudden change of $\langle \Sigma \rangle $ appears in
Fig. (\ref{fig:fig6}) when $\boldsymbol{k}_{\parallel }$ sweeps across the
anticrosssing point $\boldsymbol{k}_{a}$ so that the main characteristic of $%
\left\vert \Psi _{CB1,\pm }\right\rangle $ ($\left\vert \Psi _{VB1,\pm
}\right\rangle $) states change from hole-like (electron-like) to
electron-like (hole-like). In addition, we find sign reversals occur for $%
\langle \Sigma \rangle $ near the anticrossing points, which means the spin
orientations do not maintain the same direction. This leads to the failure
of recognizing the \textquotedblleft spin-up\textquotedblright ~and
\textquotedblleft spin-down\textquotedblright ~branches simply by their spin
orientations. Therefore we should classify the different spin states in a
new set of basis functions as discussed in Sec. \ref{sec:theoryB}.
Increasing the thickness of middle AlSb barrier, i.e., weakening of the
interlayer coupling between InAs and GaSb layers, makes the smooth variation
of $\langle \Sigma \rangle $ more and more sharp.

\begin{figure}[t]
\includegraphics[width=1\columnwidth]{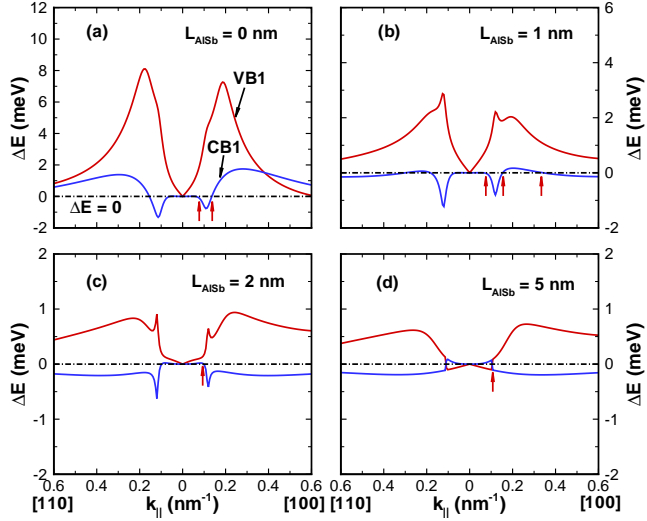}
\caption{(Colour online) Rashba spin-splitting of $CB1$ (blue line) and $VB1$
(red line) subbands in InAs/AlSb/GaSb quantum well structures with different
thicknesses of AlSb layers: (a) $L_{AlSb} = 0$ (b) $L_{AlSb} = 1$ nm (c) $%
L_{AlSb} = 2$ nm (d) $L_{AlSb} = 5$ nm. The thickness of InAs and GaSb
layers are fixed at 10 nm. The red arrows mark the cross point between the
curve of RSS and the dashdotted line $\Delta E = 0$.}
\label{fig:fig7}
\end{figure}

Besides the spin orientations in InAs/AlSb/GaSb QWs, it is interesting to
discuss the zero-field spin-splitting in these QWs because it can be
directly measured from the experiments\cite{JLuo}. Therefore, we plot the
Rashba spin-splitting (RSS) of $CB1$ and $VB1$ subbands as a function of the
in-plane momentum in Fig. \ref{fig:fig7} (a). From the figure, one can see a
valley and sign-reversal occurs in the RSS of $CB1$ subband, leading to the
oscillating behavior. This anomalous behavior arises from the difference
between the anticrossing point between the $CB1(+)$ and $VB1(+)$ subbands
and that between the $CB1(-)$ and $VB1(-)$ subbands. The decrease of RSS
appearing at large $\boldsymbol{k_{\parallel}}$ is caused by the weakening
of the conduction-valence band-coupling for carriers with large momentum,
i.e., large kinetic energy or large effective bandgap\cite{NonlinearRSS}. In
Fig. \ref{fig:fig7}, we have marked the $\Delta E=0$ (spin degeneracy)
points with red arrows. By comparing to Fig. \ref{fig:fig5}, we find these
points actually lead to the maximum hybridized points in Fig. \ref{fig:fig5}%
. The splitting of $VB1$ subband is much larger than that of $CB1$ subband.
This reflects the fact that the spin-orbit coupling in valence band is much
stronger than that of conduction band. An extremum appears in the RSS of $%
VB1 $ subband near the anticrossing point. Fig. \ref{fig:fig7} (b)-(d) shows
the RSS of QWs with a AlSb barrier inserted between InAs and GaSb layers.
When introducing an AlSb barrier into InAs/GaSb QW, the asymmetry at the
left and right interfaces for InAs and GaSb layers is compensated, so the
RSS of InAs/AlSb/GaSb QWs decreases greatly with increasing the thickness of
AlSb layer. The valley in RSS of $CB1$ subband becomes sharper because the
anticrossing behavior between the $CB1$ and $VB1$ subbands is heavily
weakened as the AlSb barrier thickness increases. Interestingly, the
anticrossing behavior seems like a crossing for the thick middle AlSb
barrier, e.g., $L_{AlSb}=5$nm.

\begin{figure}[t]
\includegraphics[width=1\columnwidth]{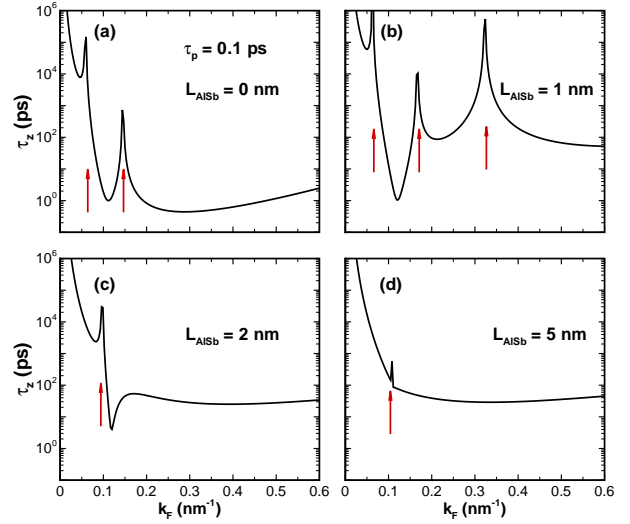}
\caption{(Colour online) Dyakonov-Perel spin relaxation time of $CB1$
subband as a function of Fermi wave vector $k_{F}$ in InAs/AlSb/GaSb QWs
with different thicknesses of AlSb layers: (a) $L_{AlSb} = 0$ (b) $L_{AlSb}
= 1$ nm (c) $L_{AlSb} = 2$ nm (d) $L_{AlSb} = 5$ nm. The red arrows mark the
resonant peeks corresponding to the spin-degeneracy points.}
\label{fig:fig8}
\end{figure}

Since the spin-splitting and spin states in InAs/GaSb and InAs/AlSb/GaSb
broken-gap QWs are very different from that in conventional semiconductor
QWs, the spin-related properties in these QWs should manifest a distinct
feature as a consequence. As an example, the DP spin relaxation time of $CB1$
subband in InAs/GaSb and InAs/AlSb/GaSb broken-gap QWs is calculated by
taking axial approximation and based on the perturbation theory\cite%
{Averkiev}. This theory demonstrates that the DP spin relaxation rate $%
\tau_{z}^{-1} \propto \Omega^{2} \propto(\Delta E_{CB1}^{ax})^2$, where $%
\Omega$ is the spin-obit coupling induced in-plane effective magnetic field
and proportional to the spin splitting $\Delta E_{CB1}^{ax}$. The
perturbation theory gives a clear physical picture about the DP spin
relaxation, that the DP spin relaxation time would show resonant peaks when
the spin splitting vanishes. As shown in Fig. \ref{fig:fig8}, for a momentum
relaxation time $\tau_{p}=0.1$ ps, we find the DP spin relaxation time in
these QWs varies from $1$ ps to $10^{5}$ ps with different $k_{F}$, and the
DP spin relaxation time $\tau_{z}$ exhibits an obvious oscillating behavior.
The resonant peaks (marked with the red arrows), actually corresponds to the
$\Delta E=0$ in (spin degeneracy) points Fig. \ref{fig:fig7}. From panel
(a)-(d), we can see the spin relaxation time in InAs/AlSb/GaSb QWs is very
sensitive to the thicknesses of middle AlSb layer. The oscillating and
large-scale variation features of DP spin relaxation time in InAs/GaSb/AlSb
QWs are dramatically different from that in conventional semiconductor QWs.
We suppose the unique features of spin relaxation time InAs/GaSb/AlSb
broken-gap QWs could provide us an interesting way to manipulate the
evolution of electron spins .

\begin{figure}[t]
\includegraphics[width=1\columnwidth]{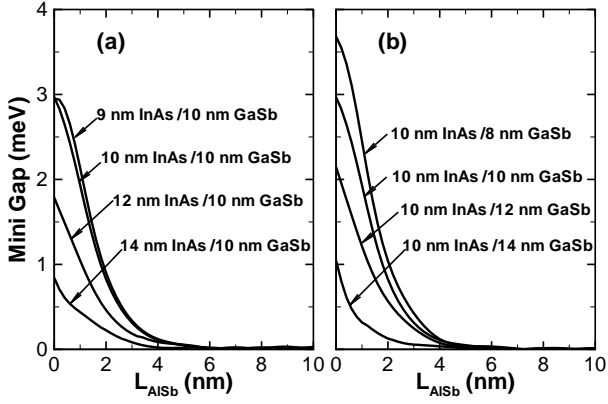} 
\caption{Mini gap as a function of the thickness of AlSb barrier in
InAs/AlSb/GaSb QWs with (a) $L_{GaSb}$ fixed at 10 nm, (b) $L_{InAs}$ fixed
at 10 nm.}
\label{fig:fig9}
\end{figure}

In Fig. \ref{fig:fig9} we show the mini gap as a function of the thickness
of the middle AlSb barrier in InAs/AlSb/GaSb QWs. The mini gap describes the
degree of electron-hole hybridization. Fix the thickness of InAs and GaSb
layer of InAs/AlSb/GaSb QW and increase the thickness of middle AlSb
barrier, the mini gap decrease rapidly, this feature was already
demonstrated experimentally in Ref. \onlinecite{MiniGap4}, in which a 1.75
meV mini gap has been measured for a 15 nm-10 nm InAs/GaSb QW and a 1.75 meV
mini gap for a 15 nm-1.5 nm-15 nm InAs/AlSb/GaSb QW. The mini gap measured
from the experiments in Refs.\onlinecite{MiniGap1, MiniGap2, MiniGap3}, are
4, 7 and 2 meV respectively. Based on our calculation, the mini gap is found
to be 0$\sim $4 meV, the order agree with the experiments. In general, we
can see the mini gaps reduce to zero when the thickness of the AlSb barrier
become larger than 5 nm. This is because the tunneling between InAs
conduction band and GaSb valence band is greatly suppressed by the AlSb
barrier, so the electron-hole hybridization is restricted. In addition, if
we fix AlSb barrier, and increase the thickness of InAs layer or AlSb layer,
the mini gap decrease too. This is because the confining energy is reduced
as the thickness of InAs layer or AlSb layers increases, and the
anticrossing point is then moved towards a higher $\boldsymbol{k}_{\parallel
}$, which has less conduction-valence interband coupling strength and forms
a smaller mini gap. As the mini gap and the hybridization degree change, the
spin-related properties, including the spin orientations, spin splitting,
and DP spin relaxation time change consequently. Therefore we certainly find
a method to tune the spin states in InAs/AlSb/GaSb QWs, which might be taken
advantages in designing spintronic devices.

\section{Conclusion}

We have investigated theoretically the spin orientation, spin-splitting,
spin relaxation in InAs/AlSb/GaSb broken-gap QWs. We found the spin states
in these broken-gap QWs are very different from that in conventional
semiconductor QWs. The spin orientations deviate away from the tangent
direction of the energy constant surface and the RSS of the anticrossed $CB1$
subband in InAs/AlSb/GaSb QWs exhibits a nonlinear and oscillating behavior.
The deviation of spin orientation comes from the strong hybridization
between different spin states, and the oscillating behavior of RSS is a
result of the anticrossing of energy dispersions according. The distinct
properties of RSS in InAs/AlSb/GaSb QWs lead to an oscillating behavior of
DP spin relaxation time obtained from the perturbation theory. By changing
the thickness of AlSb barrier between the InAs and GaSb layers, the degree
of hybridization can be tuned heavily, the mini hybridization gap and spin
orientations on the Fermi surface can be changed as a consequence. Our
theoretical calculation is interesting both from the basic physics and
potential application of the spintronic devices based on this novel
broken-gap semiconductor QW system.

\begin{acknowledgments}
This work was supported by the NSFC Grant Nos. 60525405 and 10874175, , and
bilateral program between Sweden and China.
\end{acknowledgments}

\begin{appendix}
\section{eight-band Spin Matrices}\label{appen:SM}
Following the derivation of R. Winkler\cite{ElectronSpin}, we can obtain the form of eight-band spin matrices vector in the basis
set (\ref{eqn:basis}) by $\mathbf{\Sigma'}=\mathbf{\sigma} \otimes
\mathbb{L}_{orb}$, where
$\mathbf{\sigma}=(\sigma_x,\sigma_y,\sigma_z)$ is the vector of
Pauli spin matrices, and $\mathbb{L}_{orb}$ refers to the orbital
part of the set basis function (\ref{eqn:basis}). The components
$\Sigma_x', \Sigma_y', \Sigma_z'$ of $\mathbf{\Sigma'}$ can be
written as:
\begin{equation}\label{eq:SMx}
\Sigma_{x}'=
\begin{bmatrix}
0 & 1 & 0 & 0 & 0 & 0 & 0 & 0\\
1 & 0 & 0 & 0 & 0 & 0 & 0 & 0\\
0 & 0 & 0 & \frac{i}{\sqrt{3}} & 0 & 0 & \frac{\sqrt{6}}{3} & 0\\
0 & 0 & -\frac{i}{\sqrt{3}} & 0 & \frac{2}{3}i & 0 & 0 & -\frac{\sqrt{2}}{3}\\
0 & 0 & 0 & -\frac{2}{3}i & 0 & \frac{i}{\sqrt{3}} & \frac{\sqrt{2}}{3} & 0\\
0 & 0 & 0 & 0 & -\frac{i}{\sqrt{3}} & 0 & 0 & -\frac{\sqrt{6}}{3}\\
0 & 0 & \frac{\sqrt{6}}{3} & 0 & \frac{\sqrt{2}}{3} & 0 & 0 & \frac{1}{3}i\\
0 & 0 & 0 & -\frac{\sqrt{2}}{3} & 0 & -\frac{\sqrt{6}}{3} &
-\frac{1}{3}i & 0
\end{bmatrix}
\end{equation}
\begin{equation}\label{eq:SMy}
\Sigma_{y}'=%
\begin{bmatrix}
0 & -i & 0 & 0 & 0 & 0 & 0 & 0\\
i & 0 & 0 & 0 & 0 & 0 & 0 & 0\\
0 & 0 & 0 & \frac{1}{\sqrt{3}} & 0 & 0 & -i\frac{\sqrt{6}}{3} & 0\\
0 & 0 & \frac{1}{\sqrt{3}} & 0 & \frac{2}{3} & 0 & 0 & i\frac{\sqrt{2}}{3}\\
0 & 0 & 0 & \frac{2}{3} & 0 & \frac{1}{\sqrt{3}} & i\frac{\sqrt{2}}{3} & 0\\
0 & 0 & 0 & 0 & \frac{1}{\sqrt{3}} & 0 & 0 & -i\frac{\sqrt{6}}{3}\\
0 & 0 & \frac{\sqrt{6}}{3}i & 0 & -\frac{\sqrt{2}}{3}i & 0 & 0 & \frac{1}{3}\\
0 & 0 & 0 & -\frac{\sqrt{2}}{3}i & 0 & \frac{\sqrt{6}}{3}i &
\frac{1}{3} & 0
\end{bmatrix}
\end{equation}
\begin{equation}\label{eq:SMz}
\Sigma_{z}'=
\begin{bmatrix}
1 & 0 & 0 & 0 & 0 & 0 & 0 & 0\\
0 & -1 & 0 & 0 & 0 & 0 & 0 & 0\\
0 & 0 & 1 & 0 & 0 & 0 & 0 & 0\\
0 & 0 & 0 & \frac{1}{3} & 0 & 0 & i\frac{2\sqrt{2}}{3} & 0\\
0 & 0 & 0 & 0 & -\frac{1}{3} & 0 & 0 & -i\frac{2\sqrt{2}}{3}\\
0 & 0 & 0 & 0 & 0 & -1 & 0 & 0\\
0 & 0 & 0 & -\frac{2\sqrt{2}}{3}i & 0 & 0 & -\frac{1}{3} & 0\\
0 & 0 & 0 & 0 & \frac{2\sqrt{2}}{3}i & 0 & 0 & \frac{1}{3}%
\end{bmatrix}
\end{equation}
Note that the set of basis functions we used are a little different
from the basis used in Ref. \onlinecite{ElectronSpin}, so the form
of spin matrices (\ref{eq:SMx})-(\ref{eq:SMz}) are different from
these in Ref. \onlinecite{ElectronSpin}. Using Eq.
(\ref{eqn:newbasis}), we can transform $\Sigma_x', \Sigma_y',
\Sigma_z'$ into $\Sigma_x(\varphi), \Sigma_y(\varphi),
\Sigma_z(\varphi)$, which are the components of eight-band spin
matrices in the new basis set. $\Sigma_x(\varphi),
\Sigma_y(\varphi), \Sigma_z(\varphi)$ can be written as
\begin{eqnarray}\label{Sx}  
\Sigma_{x}(\varphi) =
\begin{bmatrix}
\mathcal{B} \sin{\varphi} & i \mathcal{C} \cos{\varphi} \\
-i \mathcal{C}^T \cos{\varphi} & \mathcal{D} \sin{\varphi}
\end{bmatrix}, 
\end{eqnarray}
\begin{eqnarray}\label{Sy}
\Sigma_{y}(\varphi) =
\begin{bmatrix}
-\mathcal{B} \cos{\varphi} & i \mathcal{C} \sin{\varphi} \\
-i \mathcal{C}^T \sin{\varphi} & - \mathcal{D} \cos{\varphi}
\end{bmatrix},
\end{eqnarray}
\begin{eqnarray}\label{Sz}
\Sigma_{z}(\varphi) =
\begin{bmatrix}
0 & \mathcal{F} \\
\mathcal{F} & 0
\end{bmatrix}, 
\end{eqnarray}
where $\mathcal{B}$, $\mathcal{C}$, $\mathcal{D}$, $\mathcal{F}$ are
$4\times4$ matrices
\begin{eqnarray}\label{eq:BCDF}
\mathcal{B} & =
\begin{bmatrix}
-1 & 0 & 0 & 0 \\
0 & 0 & -\frac{1}{\sqrt{3}} & -\frac{\sqrt{6}}{3} \\
0 & -\frac{1}{\sqrt{3}} & -\frac{2}{3} & \frac{\sqrt{2}}{3} \\
0 & -\frac{\sqrt{6}}{3} & \frac{\sqrt{2}}{3} & -\frac{1}{3}
\end{bmatrix},~
\mathcal{C} & =
\begin{bmatrix}
-1 & 0 & 0 & 0 \\
0 & 0 & \frac{1}{\sqrt{3}} & \frac{\sqrt{6}}{3} \\
0 & -\frac{1}{\sqrt{3}} & -\frac{2}{3} & \frac{\sqrt{2}}{3} \\
0 & -\frac{\sqrt{6}}{3} & \frac{\sqrt{2}}{3} & -\frac{1}{3}
\end{bmatrix}, \nonumber \\
\mathcal{D} & =
\begin{bmatrix}
1 & 0 & 0 & 0 \\
0 & 0 & -\frac{1}{\sqrt{3}} & -\frac{\sqrt{6}}{3} \\
0 & -\frac{1}{\sqrt{3}} & \frac{2}{3} & -\frac{\sqrt{2}}{3} \\
0 & -\frac{\sqrt{6}}{3} & -\frac{\sqrt{2}}{3} & \frac{1}{3}
\end{bmatrix},~
\mathcal{F} & =
\begin{bmatrix}
1 & 0 & 0 & 0 \\
0 & 1 & 0 & 0 \\
0 & 0 & \frac{1}{3} & -\frac{2\sqrt{2}}{3} \\
0 & 0 & -\frac{2\sqrt{2}}{3} & -\frac{1}{3}
\end{bmatrix}. \nonumber \\
\end{eqnarray}
For a pure \textquotedblleft spin-down\textquotedblright~state
$\left\vert \Psi_{s,-}(k_{\parallel})\right\rangle $, the
expectation value of $\Sigma_x(\varphi), \Sigma_y(\varphi),
\Sigma_z(\varphi)$ can be evaluated by
\begin{subequations}
\begin{align}
\langle\Sigma_x(\boldsymbol{k}_\parallel)\rangle_{s,-} = &
\left\langle \Psi_{s,-}(k_{\parallel})\right\vert
\Sigma_x(\varphi)\left\vert \Psi_{s,-}(k_{\parallel}) \right\rangle
= \left\langle\label{eq:SOdn}
\mathcal{B}(k_{\parallel}) \right\rangle_{s,-} \sin{\varphi}, \\
\langle\Sigma_y(\boldsymbol{k}_\parallel)\rangle_{s,-} = &
\left\langle \Psi_{s,-}(k_{\parallel})\right\vert
\Sigma_y(\varphi)\left\vert \Psi_{s,-}(k_{\parallel}) \right\rangle
= -\left\langle
\mathcal{B}(k_{\parallel}) \right\rangle_{s,-} \cos{\varphi},  \\
\langle\Sigma_z(\boldsymbol{k}_\parallel)\rangle_{s,-} = &
\left\langle \Psi_{s,-}(k_{\parallel})\right\vert
\Sigma_z(\varphi)\left\vert \Psi_{s,-}(k_{\parallel}) \right\rangle
= 0.
\end{align}
\end{subequations}
Similarly, for a pure \textquotedblleft
spin-up\textquotedblright~state $\left\vert
\Psi_{s,+}(k_{\parallel})\right\rangle $, the expectation value of
$\Sigma_x(\varphi), \Sigma_y(\varphi), \Sigma_z(\varphi)$ are
\begin{subequations}
\begin{align}
\langle\Sigma_x(\boldsymbol{k}_\parallel)\rangle_{s,+} = &
\left\langle \Psi_{s,+}(k_{\parallel})\right\vert
\Sigma_x(\varphi)\left\vert \Psi_{s,+}(k_{\parallel}) \right\rangle=
\left\langle
\mathcal{D}(k_{\parallel}) \right\rangle_{s,+} \sin{\varphi},  \\
\langle\Sigma_y(\boldsymbol{k}_\parallel)\rangle_{s,+}= &
\left\langle \Psi_{s,+}(k_{\parallel})\right\vert
\Sigma_y(\varphi)\left\vert \Psi_{s,+}(k_{\parallel}) \right\rangle
=  -\left\langle
\mathcal{D}(k_{\parallel}) \right\rangle_{s,+} \cos{\varphi}, \\
\langle\Sigma_z(\boldsymbol{k}_\parallel)\rangle_{s,+} = &
\left\langle \Psi_{s,-}(k_{\parallel})\right\vert
\Sigma_x(\varphi)\left\vert \Psi_{s,-}(k_{\parallel}) \right\rangle
= 0. \label{eq:SOup}
\end{align}
\end{subequations}
According to Eqs. (\ref{eq:SOdn})-(\ref{eq:SOup}),
one can easily find
$\boldsymbol{k}_\parallel\cdot\langle\boldsymbol{\Sigma}(\boldsymbol{k}_\parallel)\rangle_{s,\pm}=0$.
Therefore, for pure \textquotedblleft
spin-up\textquotedblright~states or \textquotedblleft
spin-up\textquotedblright~states, the spin orientations are in $xy$
plane, and  strictly perpendicular to the in-plane wave vector.
\end{appendix}


\begin{thebibliography}{99}
\bibitem{EFdenote} The exact positon of Fermi level in an undoped InAs/GaSb
QW should locate inside the mini gap\cite{Band1,EFproof}. The slight
differece beteween our calculated $E_{F}$ and the exact value comes from the
summation of $s$ in Eqs. (\ref{eq:echarge}) and (\ref{eq:hcharge}) not
running over all subbands in our calculation.
\end{thebibliography}
\end{document}